\documentclass{article}

\usepackage{ltwol2e}
\usepackage{epsf}

\def\Journal#1#2#3#4{{#1} {\bf #2}, #3 (#4)}

\def\PLB{{\em Phys. Lett.}  B}

\def\PRD{{\em Phys. Rev.} D}

\def\lsim{\:\raisebox{-0.5ex}{$\stackrel{\textstyle<}{\sim}$}\:}

\def\be{\begin{equation}}
\def\ee{\end{equation}}
\def\bea{\begin{eqnarray}}
\def\eea{\end{eqnarray}}
\begin{document}

\title{RADIATIVE FOUR NEUTRINO MASSES AND MIXINGS}

\author{PROBIR ROY}

\address{Tata Institute of Fundamental Research, Mumbai, India
\\E-mail: probir@theory.tifr.res.in}   

\twocolumn[\maketitle\abstracts{
The radiative model, proposed by Zee, is extended from the three
neutrino $(\nu_{e,\mu,\tau})$ sector to include a fourth sterile
neutrino $\nu_s$.  The oscillations $\nu_e \leftrightarrow \nu_s$
(solar), $\nu_\mu \leftrightarrow \nu_\tau$ (atmospheric) and $\nu_e
\leftrightarrow \nu_\mu$ (LSND) are invoked to explain all oscillation
data with the successful relation $(\Delta m^2)_{atm} \simeq 2[(\Delta
m^2)_{solar} (\Delta m^2)_{LSND}]^{1/2}$ as an added bonus.}]

\begin{table*}[t]
\begin{center}
\caption{Experimental constraints on neutrino properties}\label{tab:experi1}
\vspace{0.2cm}
\begin{tabular}{|l|c|c|} \hline 
\raisebox{0pt}[12pt][6pt]{Neutrino anomaly} & 
\raisebox{0pt}[12pt][6pt]{Differences of mass squared} & 
\raisebox{0pt}[12pt][6pt]{Mixing angle} \\ 
\hline
\raisebox{0pt}[12pt][6pt]{Atmospheric}  &
\raisebox{0pt}[12pt][6pt]{$\delta m^2_{\mu\beta} \sim (2-5)
\times 10^{-3}$ eV$^2$} & 
\raisebox{0pt}[12pt][6pt]{$\sin^2 2\theta_{\mu\beta} > 0.8$} \\
\hline 
\raisebox{0pt}[12pt][6pt]{LSND} & 
\raisebox{0pt}[12pt][6pt]{$\delta m^2_{e\mu} \sim (0.1-1)$ eV$^2$} &
\raisebox{0pt}[12pt][6pt]{$\sin^2 2\theta_{e\mu} = (6 \pm 3) \times
10^{-3}$} \\ 
\hline
\raisebox{0pt}[12pt][6pt]{Solar Small angle} & 
\raisebox{0pt}[12pt][6pt]{$\delta m^2_{e\alpha} = (1-10) \times
10^{-6}$ eV$^2$} & 
\raisebox{0pt}[12pt][6pt]{$\sin^2 2\theta_{e\alpha} \simeq 5 \times
10^{-3}$} \\ 
\raisebox{0pt}[12pt][6pt]{MSW solution} & & \\
\hline
\raisebox{0pt}[12pt][6pt]{Solar Large angle} & 
\raisebox{0pt}[12pt][6pt]{$\delta m^2_{e\alpha} = (0.1-10) \times
10^{-5}$ eV$^2$} & 
\raisebox{0pt}[12pt][6pt]{$\sin^2 2\theta_{e\alpha} > 0.4$} \\
\raisebox{0pt}[12pt][6pt]{MSW solution} & & \\
\hline
\raisebox{0pt}[12pt][6pt]{Solar Vacuum solution}&
\raisebox{0pt}[12pt][6pt]{$\delta m^2_{e\alpha} \simeq 10^{-10}$
eV$^2$} & \raisebox{0pt}[12pt][6pt]{$\sin^2 2\theta_{e\alpha} > 0.75$} \\
\hline
\end{tabular}
\end{center}
\end{table*}

We have summarized the constraints on neutrino masses and mixings
(within two-flavor oscillations at a time) from observed neutrino
anomalies in Table \ref{tab:experi1}.  We have used the notation
$\delta m^2_{ij} 
\equiv |m^2_i - m^2_j|$ for the mass squared difference pertaining to
flavor eigenstates $i,j$ and $\theta_{ij}$ for their mixing angle.
Here $\alpha,\beta$ can be any flavor other than $e,\mu$ since $\alpha
\neq \mu$ and $\beta \neq e$ are implied by other data.

Let us first review the radiative Zee model \cite{one} which was constructed
for the $\nu_e - \nu_\mu - \nu_\tau$ system.  The electroweak gauge
group is $SU(2)_L \times U(1)_Y$.  In addition to the SM leptons and
two doublet Higgs bosons $\Phi_{1,2} \equiv \left(\matrix{\phi^+_{1,2}
\cr \phi^0_{1,2}}\right)$, with $\Phi_2$ being leptophobic, Zee also
postulated a singlet charged Higgs $\chi^+$ carrying lepton number $L
= -2$.  Thus the relevant part of the interaction is 
\bea
{\cal L}^{Zee}_1 &=& \sum_{i,j} f_{[ij]} (\nu_{i\ell} \ell_{jL} -
\ell_{iL} \nu_{jL})\chi^+ \nonumber \\ & & + \sum_i h_i(\bar\nu_{iL}
\phi^+_1 + \bar\ell_{iL} \phi^0_1)\ell_{iR} \nonumber \\ & & + \mu
(\phi^+_1 \phi^0_2 - \phi^0_1 \phi^+_2)\chi^- + h.c.
\label{eq:one}
\eea
In (\ref{eq:one}) $i,j$ are generation indices and $h_i \equiv m_{\ell
i}/\langle \phi^0_1\rangle$ are generation-hierarechical Yukawa
coupling strengths of $\Phi_1$ whereas $f_{[ij]}$, the Yukawa
couplings for the charged Higgs $\chi^+$, are antisymmetric in $i,j$;
furthermore, we have used the notation $\psi \xi \equiv \bar\psi^C
\xi$ for any two fermionic spinor fields $\psi$ and $\xi$.  The $D =
4$ terms in ${\cal L}_1^{Zee}$ above possess a Glashow-Weinberg \cite{two}
$Z_2$ symmetry, softly broken by the Higgs self-interaction
proportional to $\mu$, protecting the system against unwanted FCNC
effects.

\begin{figure}[htb]
\vskip3truecm
\includegraphics{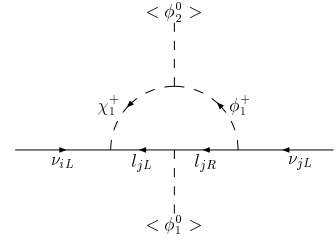}
\caption{One loop radiative $\nu_i - \nu_j$ ($i,j$ = $e,\mu,\tau$)
mass due to charged Higgs exchange.}
\label{f1}
\end{figure}

The first two terms in the RHS of (\ref{eq:one}) radiatively generate
off diagonal neutrino Majorana masses through the one-loop diagram of
Fig. \ref{f1} involving charged Higgs exchange.  The consequent
Majorana mass-matrix is \cite{three}

\begin{equation}
m_\nu \ = \ \ \matrix{& \nu_{eL} & \nu_{\mu L} & \nu_{\tau L} \cr
\overline{\nu_{eL}^{~~C}} & o & a & b \cr 
\overline{\nu_{\mu L}^{~~C}} & a & o & c \cr 
\overline{\nu_{\tau L}^{~~C}}& b & c & o},
\label{eq:two}
\end{equation}
with
\begin{equation}
 a = f_{[e \mu]} (m_\mu^2 - m_e^2) \left( {\mu v_2 \over v_1} \right)
 F(m_{\chi_1}^2, m_{\phi_1}^2), 
\label{eq:three}
\end{equation}
\begin{equation}
 b = f_{[e \tau]} (m_\tau^2 - m_e^2) \left( {\mu v_2 \over v_1} \right)
 F(m_{\chi_1}^2, m_{\phi_1}^2), 
\label{eq:four}
\end{equation}
\begin{equation}
 c = f_{[\mu \tau]} (m_\tau^2 - m_\mu^2) \left( {\mu v_2 \over v_1}
 \right)  F(m_{\chi_1}^2, m_{\phi_1}^2). 
\label{eq:five}\
\end{equation}
In (\ref{eq:three})
\begin{equation}
F(m^2_\chi,m^2_{\phi_1}) = {1 \over 16\pi^2} {1 \over m^2_\chi -
m^2_{\phi_1}} \ell n {m^2_\chi \over m^2_{\phi_1}},
\label{eq:six}
\end{equation}
with $m_{\chi,\phi_1}$ being the masses associated with $\chi^+$,
$\phi_1$.  It follows by considering the $\nu_\mu - \nu_\tau$
submatrix of (\ref{eq:two}) that the corresponding mixing angle
$\theta_{\mu\tau}$ is given by
\begin{equation}
\sin 2\theta_{\mu\tau} = {2a \over \sqrt{(b-c)^2 + 4a^2}}.
\label{eq:seven}
\end{equation}
It may also be noted that experimental constraints \cite{three} from
electron-neutrino scattering and from the decay $\tau \rightarrow \mu
\bar\nu_\mu \nu_\tau$ respectively imply 
\begin{equation}
{|f_{[e\mu]}|^2 \over M^2_\chi} \ < \ 0.036 \ G_F,
\label{eq:eight}
\end{equation}
\begin{equation}
{|f_{[\mu\tau]}|^2 \over M^2_\chi} \ < \ 0.13 \ G_F.
\label{eq:nine}
\end{equation}
Evidently, the Zee model is unable to incorporate all the constraints
of Table \ref{tab:experi1} since it does not have $\nu_s$.

\begin{table}[!ht]
\begin{center}
\caption{List of fermion and scalar fields in our model}\label{tab:experi2} 
\vspace{0.2cm}
\begin{tabular}{|c|c|c|c|}
\hline
%\multicolumn{4}{|c|}{}&\multicolumn{4}{|c|}{}\\
%\multicolumn{4}{|c|}{Higgs Sector}&\multicolumn{4}{|c|}{Lepton Sector}\\
%\multicolumn{4}{|c|}{}&\multicolumn{4}{|c|}{
%( i = generation index)}\\[2mm]
%\hline
%&&&\\
fermions & L-parity & $SU(2)_L \times U(1)_Y$ & $U(1)'$ \\
\hline
&&&\\
$(\nu_i, l_i)_L$ & $-$ & $(2, -1/2)$ & 0 \\
$l_{iR}$ & $-$ & $(1,-1)$ & 0 \\
$\nu_{sL}$ & $-$ & (1,0) & 1 \\
$N_R$ & + & (1,0) & 1 \\
$S_R$ & + & (1,0) & 0 \\
&&&\\
\hline
%&&&\\
scalars & L-parity & $SU(2)_L \times U(1)_Y$ & $U(1)'$ \\
\hline
&&&\\
$(\phi_{1,2}^+,\phi_{1,2}^0)$ & + & (2,1/2) & 0 \\
$\chi_1^+$ & + & (1,1) & 0 \\
$\chi_2^+$ & + & (1,1) & 1 \\
$\chi_2^0$ & + & (1,0) & 1 \\
&&&\\
\hline
\end{tabular}
\end{center}
\end{table}

In our model \cite{four} the gauge group is $SU(2)_L \times U(1)_Y
\times U'(1)$ together with a $Z_2$ discrete symmetry which we call
$L$-parity.  The complete set of fields, along with their
transformation properties, are listed in Table \ref{tab:experi2}.  We
have included a 
fourth sterile lefthanded neutrino $\nu_{sL}$ which is massless at the
tree level.  The righthanded neutral leptons $N_R$ and $S_R$ are taken
to be heavy.  Among the Higgs scalars $\{\chi\}$, $\chi^+_1$ is the
equivalent of Zee's $\chi^+$ while $\chi^+_2$ and $\chi^0_2$ are
extra.  The doublets $\Phi_{1,2}$ are as in the Zee scenario.  The
relevant part of the interaction now is
\bea
{\cal L}_I = {\cal L}^{Zee}_I &+& \sum_i f'_i \bar\nu_{sL} \ell_{iR}
\chi^+_2 + \mu' \chi^+_1 \chi^-_2 \chi^0_2 \nonumber \\ &+& h' N_R S_R
\chi^{0^\star}_2 + h.c.
\label{eq:ten}
\eea
\begin{figure}[htb]
\vskip3truecm
\includegraphics{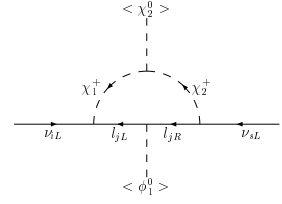}
\caption{One loop radiative $\nu_i- \nu_s$ ($i = e, \mu, \tau$)
mass due to charged Higgs exchange.}
\label{f2}
\end{figure}

Because of (\ref{eq:ten}), an extra one-loop diagram
(Fig. \ref{f2}, in addition to 
Fig. \ref{f1}, contributes to the neutrino Majorana mass
matrix $m_\nu$ of (\ref{eq:two}).  The latter is now extended to 
\begin{equation}
m_\nu \ = \ \ \matrix{&\nu_{eL} & \nu_{\mu L} &
\nu_{\tau L} & \nu_{sL} \cr 
\overline{\nu^{~~C}_{\mu L}} & o & a & b & d \cr 
\overline{\nu^{~~C}_{eL}} & a & o & c & e \cr 
\overline{\nu^{~~C}_{\tau L}} & b & c & o & f \cr
\overline{\nu^{~~C}_{sL}} & d & e & f & o}
\label{eq:eleven}
\end{equation}
Here the extra elements $d,e,f$ are given by
\begin{equation}
d = \left(f_{[e\tau]} f'_\tau m_\tau + f_{[e\mu]} f'_\mu m_\mu\right)
\mu' \langle \chi^0_2\rangle F(m^2_{\chi_1},m^2_{\chi_2}),
\label{eq:twelve}
\end{equation}
\begin{equation}
e = \left(f_{[\mu\tau]} f'_\tau m_\tau + f_{[\mu e]} f'_e m_e\right)
\mu' \langle \chi^0_2\rangle F(m^2_{\chi_1},m^2_{\chi_2}),
\label{eq:thirteen}
\end{equation}
\begin{equation}
f = \left(f_{[\tau\mu]} f'_\mu m_\mu + f_{[\tau e]} f'_e m_e\right)
\mu' \langle \chi^0_2\rangle F(m^2_{\chi_1},m^2_{\chi_2}).
\label{eq:fourteen}
\end{equation}
Neglecting $m^2_e$ in comparison with $m^2_\tau$, (\ref{eq:three}) and
(\ref{eq:twelve}) imply that 
\begin{equation}
d = {be \over c} \left(1 - {m^2_\mu \over m^2_\tau}\right) + {f
f_{[e\mu]} \over f_{[\tau\mu]}}.
\label{eq:fifteen}
\end{equation}

We assume hierarchical couplings for the lepton nonconserving Yukawa
terms, i.e. 
\begin{equation}
|f_{[e\tau]}| \ll |f_{[e\mu]}| \ll |f_{[\mu\tau]}|.
\label{eq:sixteen}
\end{equation}
This means that $|a| \ll |b| \ll |c|$.  Moreover, we assume that
$|f'_i| \lsim m_i/v_1$.  This leads us to conclude that $c$ is the
dominant matrix element in (\ref{eq:eleven}) and, if
$|f_{[\mu\tau]}|^2 \gg |f_{[e\mu]} f_{[e\tau]}|$, $|ef| \gg |ab|$.
Under these circumstances, the leading expressions for the mass
eigenvalues of (\ref{eq:eleven}) are:
\bea
m_1 &\simeq& -2 {ab \over c}, \ m_2 \simeq -2 {ef \over c},
\nonumber \\ 
m_3 &\simeq& c + {ef \over c}, \ m_4 \simeq -c + {ef \over c}.
\label{eq:seventeen}
\eea
Thus $\nu_3$ and $\nu_4$ make a pseudo-Dirac pair out of a maximally
mixed combination of $\nu_\mu$ and $\nu_\tau$.  On the other hand,
$\nu_1$ and $\nu_2$ are predominantly $\nu_e$, $\nu_s$ combinations 
with a small mixing angle. 

We now have
\bea
(\Delta m^2)_{solar} &\simeq& {4e^2 f^2 \over c^2}, \ (\Delta
m^2)_{atmos} \simeq 4ef, \nonumber \\ (\Delta m^2)_{LSND} &\simeq& c^2.
\label{eq:eighteen}
\eea
and
\[
(\Delta m^2)_{atm} \simeq 2\left[(\Delta m^2)_{solar} (\Delta
m^2)_{LSND}\right]^{1/2}. 
\]
The mixings can be summarized as
\bea
\sin^2 2\theta_{e\mu} &\simeq& 4b^2/c^2, \ \sin^2 2\theta_{\mu\tau}
\simeq 1, \nonumber \\ \sin \theta_{es} &=& {f_{[e\mu]} c \over 2
[f_{\mu\tau}] e} + {b \over 2f} {m^2_\mu \over m^2_\tau}.
\label{eq:nineteen}
\eea
A suitable parametric choice is: $a \simeq 3 \times 10^{-5} eV$, $c
\simeq 1 eV$, $e \simeq 0.12 eV$ and $f \simeq 0.01 eV$.  These yield
$m_{\nu_e} \simeq 2 \times 10^{-6} eV$, $m_{v_s} \simeq 2.4 \times
10^{-3} eV$ and $\sin^2 2\theta_{es} \sim 6 \times 10^{-3}$ to match
with the small angle MSW solution of the solar neutrino anomaly in
Table \ref{tab:experi1}.  In comparison, for the atmospheric anomaly,
we have $\Delta 
m^2_{\mu\tau} \sim 5 \times 10^{-3} eV^2$, $\sin^2 2\theta_{\mu\tau}
\simeq 1$ with $\nu_\mu$ and $\nu_\tau$ masses around $1 eV$.
Finally, for the LSND anomaly, we have $\Delta m^2_{e\mu} \sim 1 eV^2$
and $\sin^2 2\theta_{e\mu} \simeq 6 \times 10^{-3}$. 

This work has been done in collaboration with N. Gaur, A. Ghosal and
E. Ma.

\end{document}